# Long-Range QKD without Trusted Nodes is Not Possible with Current Technology

**Authors:**
**Bruno Huttner**, ID Quantique, Switzerland[†];
**Romain Alléaume**, Telecom Paris - Institut Polytechnique de Paris, France;
**Eleni Diamanti**, Sorbonne University, CNRS - LIP6, France;
**Florian Fröwis**, ID Quantique Europe, Austria;
**Philippe Grangier**, Université Paris-Saclay, IOGS, CNRS, France;
**Hannes Hübel,** Austrian Institute of Technology, Austria;
**Vicente Martin**, Center for Computational Simulation / ETSIInf. Universidad Politécnica de Madrid, Spain;
**Andreas Poppe**, Austrian Institute of Technology, Austria;
**Joshua A. Slater**, QuTech - Delft University of Technology, The Netherlands ;
**Tim Spiller**, University of York, UK;
**Wolfgang Tittel**,
QuTech and Kavli Institute of Nanoscience, Delft Technical University, The Netherlands; Department of Applied Physics, University of Geneva, Switzerland; Schaffhausen Institute of Technology in Geneva, Switzerland;
**Benoit Tranier**, ThalesAleniaSpace, France;
**Adrian Wonfor**, University of Cambridge, UK;
**Hugo Zbinden**, Department of Applied Physics, University of Geneva, Switzerland.

## Abstract

A recently published patent[1] has claimed the development of a novel quantum key distribution protocol purporting to achieve long-range quantum security without trusted nodes and without use of quantum repeaters. Here we present a straightforward analysis of this claim, and reach the conclusion that it is largely unfounded.

## Introduction

Today, point-to-point quantum key distribution (QKD) is already a commercial reality. The range of commercial QKD systems is typically one hundred kilometers over optical fibers. Academic systems and new protocols can reach several hundreds of kilometers[2],[3]. Free-space QKD links to Low Earth Orbit satellites have been demonstrated by the Chinese Micius satellite[4]. However, the range of a single point-to-point link is still limited by the power loss of the link[5]. To expand real-world applications of QKD it is necessary to extend the range up to global QKD and offer more complex network topologies[6]. Pending new technologies, such as the quantum repeater, this expanded versatility is achieved by so-called Trusted Nodes (TNs)[7]. In a TN, the quantum signal is measured and converted to a classical signal. A new classical signal is generated, converted to quantum, and sent to the next node. TNs can be used as relays, to provide long-distance QKD, and as switches, to provide complex topologies[6]. However, since the TN contains a classical signal, which can in principle be copied, there is no quantum security within the TN. The TN must be trusted and physically protected, to avoid any leakage of the data[6]. Therefore, for security purposes, TNs represent weak points in a complete end-to-end QKD transmission. In this paper, the term long-distance QKD means global QKD, i.e. the ability to deploy and implement QKD between any two points on Earth.

---

[†] Corresponding author: bruno.huttner@idquantique.com

Recently, the UK Intellectual Property Office granted the patent number GB2590064[1] to the company Arqit Ltd. In the following, we will refer to this patent as the ARQ19 patent. We will also refer to the protocol described in this patent as the ARQ19 protocol. This patent purports to offer long-distance QKD with no trusted nodes. According to the claims, global QKD could be achieved now with untrusted satellites. This would represent a game-changer for QKD. Therefore, it is clearly important that these claims are investigated. Unfortunately, as far as we are aware, they have not been validated by an accompanying public disclosure in any scientific journal. Therefore, we base our analysis on the ARQ19 patent and on the 20-F annual report filed by Arqit at the Securities and Exchange Commission (SEC)[8]. This report will be referred to as the SEC filing. We present here our results. We believe that an open discussion is essential to the provision of trustful solutions for future secure communications, quantum or otherwise. This is similar to the process currently being carried out by the National Institute of Science and Technology (NIST) for Post-Quantum Cryptography. In this process, the NIST has solicited researchers worldwide to submit quantum-resistant public-key cryptographic algorithms. The NIST has been evaluating them and has now decided to standardize some of them. To improve trust in the choice, all the candidate algorithms were presented and discussed publicly[9].

## Quick Reminder on QKD

In the following, we follow the convention of the ARQ19 patent: the two end-users are named Bob and Carol. The TN, which acts as a broker for Bob and Carol is named Alice. The two partners, Bob and Carol, want to exchange confidential information. Quantum Key Distribution (QKD) provides a solution to establish a shared secret key with information-theoretic security[3],[10]. QKD guarantees that the key cannot be broken in the future, even by a computationally unbounded adversary. The secret key established by QKD can then be combined with an authenticated encryption scheme to perform secure communication between Bob and Carol, with a security gain compared to conventional key distribution methods[10]. In particular, secure communication with information-theoretic security can be obtained by combining QKD with One-Time-Pad encryption. Alternatively, QKD can be combined with symmetric encryption such as AES to strengthen the resilience of secure communications. In this document, we adopt the terminology *Quantum Security* for the security that can be reached in the exchange of confidential information between Bob and Carol, using QKD or any similar quantum technology providing information-theoretic security for the key exchange.

QKD relies on two elements:

1. A quantum channel, which permits the exchange of quantum objects, typically single photons, over an optical fiber (ground applications) or over free space (satellites). The main property of this quantum channel is that any attempt at eavesdropping will modify the state of the objects exchanged through the channel. This will be discovered during subsequent information reconciliation steps.
2. A classical discussion or reconciliation channel, which allows the processing of the quantum exchange and the establishment of the secret key. This channel is authenticated (this means that Bob and Carol know that they are talking to one another, and that their exchange cannot be modified). However, in standard QKD protocols, this channel is public: the information exchanged is available to all. QKD is therefore a solution for transforming a non-confidential authenticated channel into a confidential authenticated one.

The main limitation of QKD today is that the range of a quantum channel is limited. Typically, owing to unavoidable loss, a ground-based quantum channel built with optical fibers has a maximum reach of a few 100 kms. Free-space quantum channels, especially from satellites to ground, can achieve longer ranges (as most of the transmission is in the vacuum, with no optical absorption). This can be extended to a global range by means of trusted satellites. In the remaining, we shall denote as *long-range QKD* the ability to establish keys with quantum security, over any distance.

## QKD with Satellites

In the present discussion, we assume that Bob and Carol are located far apart from each other. Therefore, they need to rely on a satellite, referred to as Alice. In the standard Trusted Node model, Bob exchanges quantum objects with the satellite, then processes the exchange over the classical discussion channel to generate a key. Later on, the orbit of the satellite will take it over to Carol, and the same process is implemented. This is explained in Figure 1. By performing some more processing of the two keys, typically a bit-by-bit XOR of the two keys, the satellite, Alice, can help generate a joint secret key, known by Bob and Carol. This key can be used for encrypting their data and providing Quantum Security. The caveat here is that Alice knows the two intermediate keys, and therefore the final secret key. Therefore, it must be trusted by the users. Long-range QKD with a Low Earth Orbit trusted satellite has been demonstrated a few years ago by a Chinese team[4].

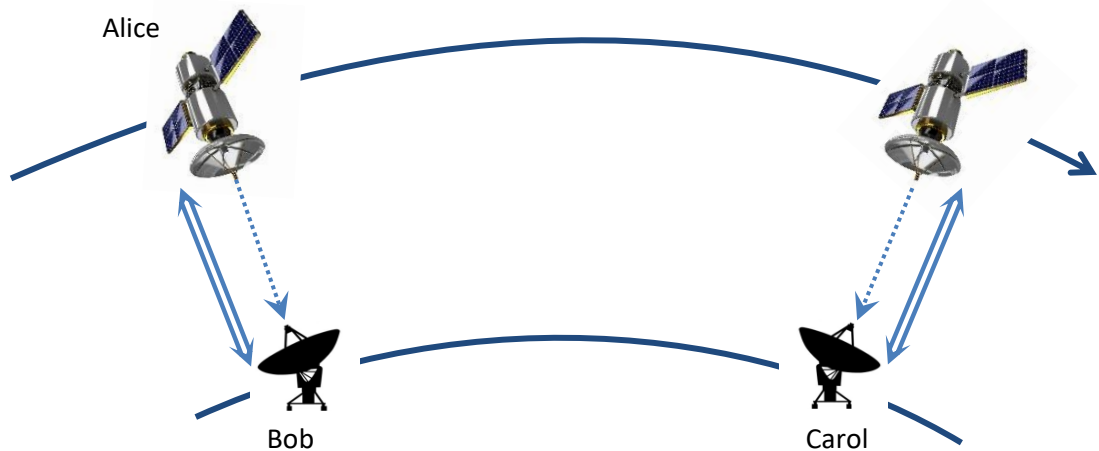

*Figure 1 Long-range quantum key exchange with satellite as Trusted Node*

*The dotted arrows represent quantum channels between the satellite (Alice) and each of the users (Bob and Carol). The white arrows represent classical public discussion channels, used to process the keys. By using both quantum channels and discussion channels, Bob and Carol can build a secret shared key. However, this secret key is also known to Alice, who therefore must be trusted by the parties.*

Having to trust an intermediary to transfer secret keys is clearly a handicap for long-range QKD. One existing solution is to use so-called entangled states. The satellite distributes simultaneously two quantum particles, one to each of the users. Entanglement ensures that the satellite knows nothing about the state of each particle, while allowing Bob and Carol to extract a secret key. Unfortunately, the fact that the two particles must be generated simultaneously requires that the users are both in the line-of-sight of the satellite, which limits the accessible range on the surface of the earth to typically 1000 km. This limitation might be eliminated in the future, through the use of quantum memories, which can store one of the two particles and release it when the satellite is above the second target. However, quantum memories of sufficient technological maturity are not yet available. Therefore, with current technology, full security by quantum key exchange with an untrusted satellite is only available for a restricted distance range; long-distance security requires trusting the satellite.

## The approach of Arqit Ltd.

The now often-repeated claim of Arqit regarding the ARQ19 protocol[1], is that it provides a new solution, which uses only regular quantum channels (no entangled states, no quantum memories) and allows the distribution of secret keys with an untrusted satellite over any range. See, for example in the SEC filing[8]: "Arqit's invention — the ARQ19 algorithm — permits random numbers to be sent globally and in a provably secure manner. Using Arqit's proprietary technology, the satellite sends random numbers which OGRs (Optical Ground Receivers) use to create keys in a manner which is entirely isolated from the satellite or any other infrastructure. Thus, the only entities that can ever know the keys created are the two OGRs involved in doing so. This means that the ARQ19 protocol is entirely secure against third parties, which is a novel class of cryptographic service."

This is a very strong claim, which has been eluding the quantum cryptography community for decades. Part of the claimed solution is the ARQ19 protocol presented in the ARQ19 patent, which has recently become public. The following discussion is therefore based on this patent.

We note that, in the above-mentioned SEC filing[8] and elsewhere, the claim is that the new method is not based on QKD, but on a new concept called Quantum Key Infrastructure or QKI. The concept of QKI was already used much earlier in a different context, in order to solve the authentication issues of QKD[11]. It was also introduced again recently, for another key distribution protocol independent of QKD[12]. Both are different from the QKI concept introduced in the SEC filing, which we analyse further. In the SEC filing, it is asserted that the technology used is not QKD, but that the new quantum algorithm solves all known problems with QKD and replaces it with this new QKI concept. It is stated that the system does not distribute keys, it distributes quantum random numbers which are an input into a key creation process involving other areas of proprietary classical cryptography.

This supposed distinction is in fact a misrepresentation of QKD. As is clear from Table 1 for example, QKD does not distribute keys directly. It distributes quantum objects, which are measured and processed jointly by the users, over the discussion channel. After this discussion, when the process is successful, the users share a secret string of random numbers, which can be used as a secret key, or for any other purpose. This is exactly what is claimed of QKI. There is thus no real distinction between QKD and QKI. In addition, the ARQ19 protocol is explicitly a QKD protocol, as mentioned in the very title of the ARQ19 patent. However, since the patent is rather lengthy (125 pages) and cumbersome, it is not easy to extract the key ideas. Therefore, we summarize the protocol in Table 1 below. For definiteness, we use the standard BB84 protocol, with polarisation encoding. Any other prepare & measure protocol could be applied in a similar way.

| | Bob | | Alice | | Carol |
|---|---|---|---|---|---|
| 1 | 1 1 X 1 X 0 0 1 X 1 0 0 X 1 | ⇜ | 1 0 1 1 1 0 0 1 0 1 1 0 1 0 | | |
| 2 | 1 1 1 0 0 1 1 0 0 1 | ⇒ | 1 0 1 0 0 1 1 1 0 0 | | |
| 3 | 1 X X 0 X 1 1 X 0 X | ⇐ | Choice of bases | | |
| 4 | | | 1 1 0 0 1 0 1 1 1 0 0 1 0 0 | ⇝ | 1 X 1 X 1 0 0 X 0 1 1 0 X 0 |
| 5 | | | 1 0 1 0 1 1 0 0 1 0 | ⇐ | 1 1 1 0 0 0 1 1 0 0 |
| 6 | | | Choice of bases | ⇒ | 1 X 1 0 X 0 X 1 X 0 |
| 7 | | | 1 0 1 0 0 1 1 1 0 0 <br> ⊕ <br> 1 0 1 0 1 1 0 0 1 0 <br> 0 0 0 0 1 0 1 1 1 0 | ⇒ | 0 0 0 0 1 0 1 1 1 0 |
| 8 | 1 X X 0 X 1 1 X 0 X | ⟶ | | | ? X X ? X ? ? X ? X |
| 9 | | | | | 0 X X 0 X 0 1 X 1 X |
| 10 | | | | | 1 X 1 0 X 0 X 1 X 0 <br> ⊕ <br> 0 X X 0 X 0 1 X 1 X <br> 1 X X 0 X 0 X X X X |
|  | ? X X ? X ? X X X X | ⟵ | | | |
| 11 | 1 X X 0 X 1 X X X X | | Shared sifted key between Bob and Carol | | 1 X X 0 X 0 X X X X |

*Table 1 : Summary of the ARQ19 protocol*

The different steps of the protocol are described below.

1. Alice prepares a series of quantum states, according to BB84 polarisation protocol. For each state, she chooses both the bit value and the corresponding basis. She sends the states to Bob over a quantum channel (arrow with diagonal stripes).

2. Many states are lost in the transmission. Bob tells Alice, which states have been lost (X in the table). He uses the classical discussion channel (white arrow). Alice and Bob discard all the corresponding states. The resulting series of bits is the raw key.

3. Alice tells Bob, over the classical discussion channel, which bases she used. Bob notes the cases when he and Alice used different bases (X in the table) but does not tell Alice. The remaining bits represent the sifted key for Bob. Alice cannot know, which of the states were received by Bob in the correct basis.

4. to 6. Alice and Carol follow the same protocol with a new series of states.

7. Alice performs an XOR of the two raw keys she exchanged with Bob and with Carol and sends the result to Carol, over the classical discussion channel.

8. Bob sends directly to Carol, which bits he received in the wrong basis and should not be used (X in the table). He uses a confidential discussion channel, which cannot be eavesdropped by Alice (black arrow).

9. Carol notes the wrong bits in the XORed key.

10. Carol makes an XOR of the two sifted keys, and sends to Bob, which bits should not be used (X in the table). She also uses the same confidential discussion channel, which cannot be eavesdropped by Alice.

11. Bob and Carol now share a common sifted key, unknown to Alice. They can process it in the standard way (error estimation, error correction, privacy amplification) to finally get a shared secret key. The main hypothesis of the protocol is that, prior to the exchange, Bob and Carol share a confidential channel, which cannot be eavesdropped by Alice.

The ARQ19 protocol proceeds as follows (all the steps refer to Table 1): Bob first exchanges quantum objects with Alice but does not fully process them to get a secret key (steps 1. to 3.). After these steps, Alice possesses a raw key (she does not know which bits were kept by Bob) and Bob possesses a sifted key (all the wrong bases are tagged and will not be used). Carol then does the same thing later, when the satellite is above her (steps 4. to 6.). Alice then prepares an XOR of the two raw keys and forwards it to Carol (step 7.).  In order to build their common sifted key, Bob now tells Carol which of the bits of his own sifted key he kept (step 8.), so that Carol knows, which bits of the XORed key to keep (step 9.). Carol performs a new XOR of her own sifted key and the sifted XORed keys sent by Bob, and tells Bob, which of the bits she removed (step 10.). Bob and Carol now share a common sifted key (step 11.), which has to be processed in a standard way to extract a common secret key. Note that Bob and Carol interact directly (i.e. not via the satellite) over a direct discussion channel (black arrow in Table 1). This ensures that the satellite does not know the secret key. It follows that Bob and Carol do not need to trust the satellite, which only acts as a broker. Bob and Carol have thus managed to exchange secret keys through an untrusted satellite.

However, there is a catch. In the ARQ19 patent and in the above explanation of Table 1, it is necessary that the direct discussion channel between Bob and Carol must be kept secret from Alice. We refer to this channel as a confidential discussion channel. The requirement that Bob and Carol share a confidential discussion channel may look like a minor requirement, but it is not. An untrusted satellite means that any data available in the satellite may be sent to the ground. The simplest action for an untrusted satellite is to broadcast the key information to any eavesdropper, as outlined in Figure 2. More generally, if the satellite were to be hacked, the hacker would need to get the data to the ground in order to use it. Therefore, the confidential discussion channel between Bob and Carol needs to be kept secret from any possible attacker. In other words, it must be a true confidential channel (not only from the satellite). The way to achieve the security of this confidential channel is not discussed in the ARQ19 patent.

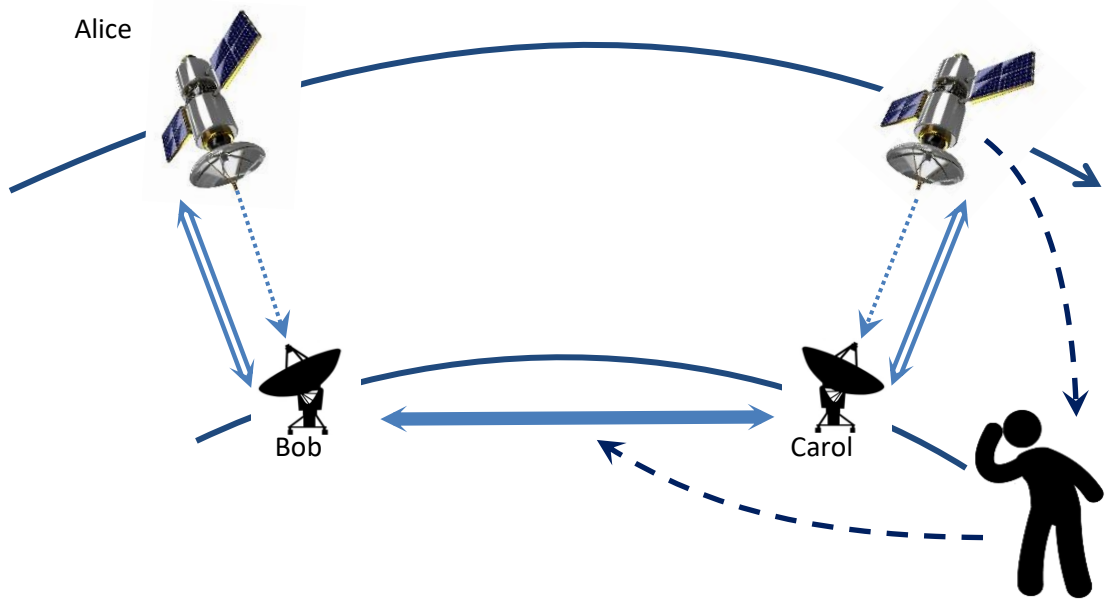

*Figure 2 : Long-range quantum key exchange with an untrusted satellite*

*In the ARQ19 protocol, the processing of the keys is only partly done through the public discussion channels with the satellite. The final processing is done directly between Bob and Carol (full arrow). However, if the satellite is untrusted, the data from the satellite can leak to an eavesdropper on the ground (dashed line). The discussion channel between Bob and Carol therefore must be protected from the eavesdropper. It must be a confidential channel.*

We are left with two alternatives. If the satellite is trusted, Bob and Carol can achieve quantum security with standard QKD. If the satellite is not trusted, the security of the whole system is only as strong as the security of the separate confidential discussion channel between Bob and Carol. Clearly, as this channel operates directly between Bob and Carol independent of their locations, it cannot be based on currently available quantum technology. Therefore, the claim that Bob and Carol can achieve quantum security with an untrusted satellite appears to be unfounded.

The above does not mean that the application of the above protocol has no interest whatsoever. Indeed, with this proposed solution, any attacker must both hack into the satellite to get the data (or equip it with unreliable devices), and break the confidential discussion channel. This means that the solution could add one layer of security. However, as we now show, the same level of security could be obtained with a standard QKD system, which does not rely on the ARQ19 protocol.

## An Equivalent Approach

In the standard method of generating a shared secret key with a relay discussed previously, Bob exchanges a QKD key with the satellite. Carol does the same when the satellite is above her. The satellite then prepares the XOR of the two keys, which is sent to Carol over a public communication channel. Carol can then recover the first key exchanged between Bob and the satellite. However, the satellite obviously also knows the shared key.

In order to add an extra layer of security, Bob can locally generate a new key, unknown to the satellite, and forward the XOR of this key with the previous one to Carol, using the same confidential discussion channel as in the ARQ19 protocol. This would allow Carol to recover the new key, which is not known to the satellite. This simple hybrid approach hence consists in combining two secret keys established over two channels: one relying on the satellite QKD, the other relying on the confidential non-quantum channel available on the ground (as assumed in the ARQ19 patent). Breaking the security of such a simple hybrid scheme requires breaking both key establishment schemes. This is exactly equivalent to what is required to break the ARQ19 protocol. Our approach here does not use any new innovative method or protocol, but simply relies on the mixing of two different cryptographic primitives. Note that the way to achieve the confidentiality of the discussion channel is not discussed in the ARQ19 patent. What is certain is that, as the distance between the ground stations is very large, it cannot be based on a quantum channel. The claim that this scheme, or an equivalent one, can provide *Quantum Security* with untrusted satellites is again unfounded.

## Summary and Conclusions

We have investigated the claim that the ARQ19 protocol[1] can achieve long-range quantum security without trusted nodes. More specifically, the claim is that using this protocol, two distant users could exchange provably secure keys through an untrusted satellite node. As far as we are aware, this claim has not been validated by a suitable public disclosure in a scientific journal. If it were shown to be correct, this claim would represent a major step forward for quantum communications and their application worldwide. Given this, it is clearly very important – for the quantum technology community, the wider security community, and, of course, any future users – to scrutinise the claim.

Based on existing public information, and especially the published ARQ19 patent at the core of the protocol, we reach the conclusion that this claim is largely unfounded. In order to remove trust in a QKD satellite, the ARQ19 protocol simply transfers this trust to a separate secure communication channel, which cannot be quantum as it operates directly between Bob and Carol independent of their locations on earth.

Today, except when using entangled photon pairs, provably secure quantum key exchange through an intermediate node is only possible with trusted satellites. Therefore, the security of any data communication platforms that utilise the ARQ19 protocol is based on the security of a separate, unspecified, although certainly not quantum, channel. Or one has to trust the satellite.

Consequently, the whole security of the QuantumCloud™ platform described in the SEC filing[8], which rely on the assumption that you can distribute keys with *Quantum Security* globally, with an untrusted satellite, therefore appears to be at risk. As demonstrated above, the satellite can only be untrusted, if we assume that the users share an independent confidential channel, prior to their use of the QuantumCloud™. Therefore, in the case of an untrusted satellite, the security of the QuantumCloud™ is only as good as the security of this undescribed confidential channel. Clearly trust must be established in the security of any new data communication platforms. This can only be obtained by proper disclosure of the methods used, along with analysis by both the quantum and the cryptographic communities. Security by obscurity has proven disastrous in the past. It should not be used to claim high-level security today.

## Note added during the reviewing process

During the reviewing process, Arqit announced an independent review on its technology by the University of Surrey:
https://arqit-res.cloudinary.com/image/upload/v1652267532/Press/Arqit_Quantum_Inc._Announces_Independent_Assurance_Report_on_its_Technology_11_May_2022__hqqifl.pdf .
We understand that this review only covers the (classical) part of the Arqit technology that follows after the initial key-sharing. Therefore, it does not include the key exchanges with the satellite, which is the topic of our paper. Thus, this review does not modify our conclusion in any way.

## Acknowledgments

B.H., R.A., E.D., F.F., P.G., H.H. V.M., A.P., J.A.S., A.W. and H.Z. acknowledge support from the H2020-funded research project OPENQKD, Grant agreement contract number 857156, https://openqkd.eu/

# Competing Interests

B.H. and F.F. are employees of ID Quantique, Geneva and ID Quantique Europe, Vienna, respectively, which have competing interests with Arqit in developing quantum communication technologies. B.T. is an employee of Thales Alenia Space, a joint Venture which invests in satellite quantum communications. BH is the inventor of several patents, both pending and accepted, in the field of space QKD. The authors declare that there are no other competing interests.

# Author Contributions

All authors have contributed equally to the manuscript.